\documentclass[12pt, draftclsnofoot, onecolumn]{IEEEtran}
\usepackage{epsfig,makeidx,color}
\usepackage{amsmath,amssymb,bbm}
\usepackage{cite,graphicx,lipsum}
\usepackage{enumerate,balance}
\usepackage[switch,pagewise]{lineno}
\usepackage{hyperref}
\hypersetup{
        colorlinks = true,
        citecolor=blue,
}
\def\rT{{\rm T}}

\def\uB{{\mathbb B}}

\def\uE{{\mathbb E}}

\DeclareMathOperator*{\argmin}{\arg\!\min}

\def\be{ \begin{equation} }
\def\ee{ \end{equation} }
\def\bea{ \begin{eqnarray} }
\def\eea{ \end{eqnarray} }

\def\bd{{\bf d}}

\def\ba{{\bf a}}

\def\bee{{\bf e}}

\def\bone{{\bf 1}}

\def\b0{{\bf 0}}

\def\cA{{\cal A}}

\ifCLASSOPTIONonecolumn
  \newcommand{\figwidth}{0.50\columnwidth}
  
\else
  \newcommand{\figwidth}{0.80\columnwidth}
  
\fi

\title{Multiuser Offloading with Cloud Server Data}

\author{Jinho Choi\\
\thanks{The author is with
the School of Information Technology,
Deakin University, Geelong, VIC 3220, Australia
(e-mail: jinho.choi@deakin.edu.au).
This research was supported
by the Australian Government through the Australian Research
Council's Discovery Projects funding scheme (DP200100391).}}

\begin{document}

\maketitle

\begin{abstract}
Computation offloading becomes useful for users of limited computing power and mobile edge computing (MEC) can help mobile users perform their tasks more effectively. In this paper, we consider MEC when users  perform tasks with local data as well as data at cloud storage servers, since users can often keep their data at cloud storage servers or have tasks that need to use public datasets. An optimization problem to minimize the total consumed energy of users as well as a base station (BS) is formulated with a total transmission time constraint, which is solved by converting into multiple binary integer linear programming problems.
\end{abstract}
\ifCLASSOPTIONonecolumn
\baselineskip 26pt
\fi

\section{Introduction}

While user equipment (UE) such as smart phones becomes more powerful in terms of storage and computing power, 
there will be more applications that need to perform tasks with more data and computing resources. Thus, it might be necessary to
exploit the powerful computing power of cloud servers via 
computation offloading  \cite{Yang08} \cite{Kumar10}. 
In cellular systems, mobile edge computing (MEC) \cite{Barbarossa14} \cite{Mach17} \cite{Abbas18} can be employed so that  MEC servers integrated into base stations (BSs) can provide computing power, and since
multiple UEs or users co-exist in a cell, joint computation offloading for multiple users can be considered \cite{Chen16} \cite{You17}.

Multiuser MEC offloading is  extended in \cite{Zhou21} to deal with different user requirements in heterogeneous networks, and in \cite{Anajemba20} for cooperative offloading in a heterogeneous network consisting of small-cell BSs and wireless relays.
In addition, as in \cite{Choi22_OL}, a distributed approach can be employed where uploading for offloading is carried out using random access so that no coordinated radio resource allocation is necessary. 

Most computation offloading approaches assume that users have data to perform tasks \cite{Chen16} \cite{Zhou21}. However, in addition to local data at users, some users' tasks may need data at cloud servers. For example, users may use cloud storage \cite{BUYYA2009599} \cite{Wu10}. In this case, users need to receive the data stored at cloud storage when they do not choose computation offloading, which may incur a significant cost of downloading if a large amount of server data is needed (e.g., for a facial recognition task at a user, downloading a database of known faces to find a match can be stored at a server can result in a high communication cost). 
As a result, computation offloading becomes a more attractive approach,
because it not only allows users to utilize the server's computing power, but also eliminates the need to download datasets that can be expensive if the amount of the server data is large.

In this paper, we study MEC offloading when users need data at  server to perform tasks. It is noteworthy that if there is no need to use server data, it boils down to conventional MEC offloading. As a result, the problem in this paper can be seen as a generalization. We formulate an optimization problem to minimize the total consumed energy for both computation and signal transmissions  subject to a total transmission time constraint for both uplink and downlink\footnote{In conventional MEC offloading problems, only uplink transmissions are considered as no data at storage servers are to be used.}. Then, we derive an approach that converts the problem into multiple binary integer linear programming problems to find the optimal set of offloading users without an exhaustive search that requires a complexity growing exponentially with the number of users.



\section{System Model}

In this section, we present the system model for wireless multiuser offloading. It is assumed that there are $K$ users and one BS. 
Throughout the paper, we assume that the computing and data storage servers are connected to the BS with a link of high data rate. Thus, the servers and BS are assumed to be interchangeable.

\subsection{Tasks with Data at a Server}

Suppose that user $k$ has a computation task with local data of size $L_k$, which is stored at user $k$, and server data of size $B_k$, which is stored at the server. Thus, the size of the aggregated data is given by $I_k = L_k+B_k$. With a slight abuse of notation, $B_k$, $L_k$, and $I_k$ are also referred to as the server data, local data, and aggregated data of the task at user $k$, respectively.
The number of central processing unit (CPU) cycles to complete the task with $I_k$ is denoted by $C_k$. In most existing works, it is assumed that the server does not have data. Thus, $I_k = L_k$ and each user is able to complete the task without communicating with the BS. However, when $B_k > 0$, it is necessary for user $k$ to communicate with the BS.  

\begin{figure}[thb]
\begin{center}
\includegraphics[width=\figwidth]{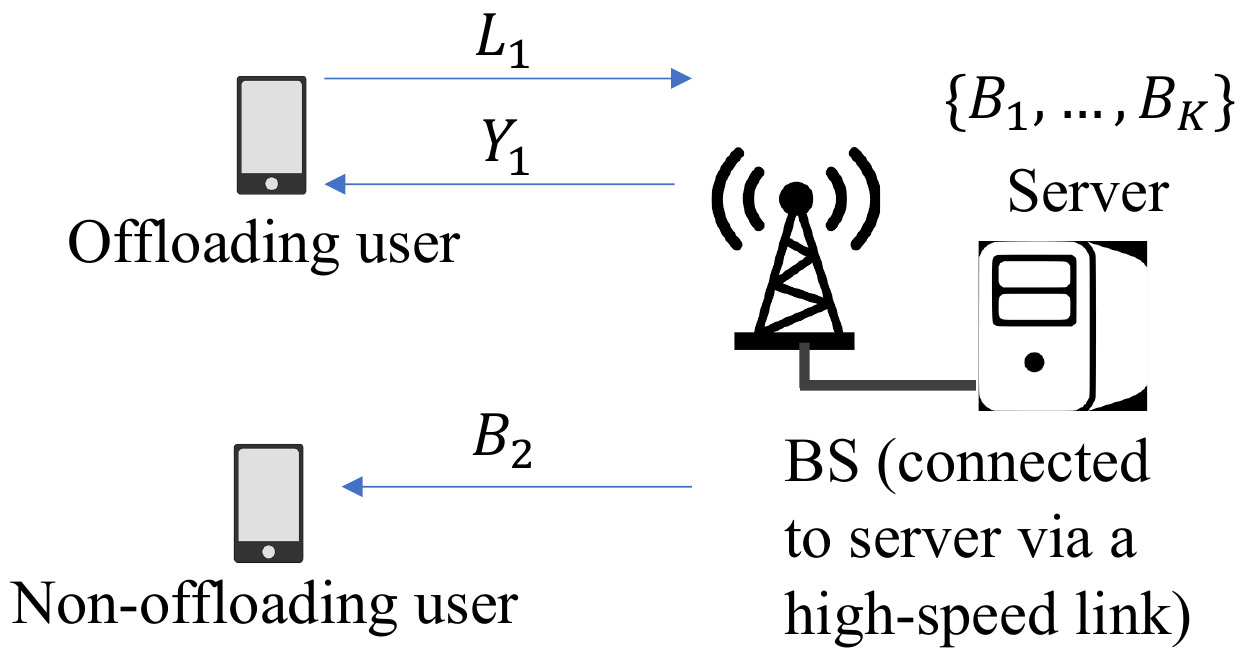} 
\end{center}
\caption{An illustration of computation offloading when users need to use datasets at the cloud storage server (here, user 1 is an offloading user and user 2 is a non-offloading user).}
        \label{Fig:Fig_DS}
\end{figure}

Let $a_k \in \{0,1\}$ denote the selection of computation offloading.  That is, if $a_k = 1$, user $k$ chooses computation offloading (this user is called an offloading user). Otherwise (i.e., $a_k =0$), user $k$ performs local computing (this user is called a non-offloading user). In addition, let 
$\cA = \{k: \ a_k = 1, \ k = 1,\ldots, K\}$
and $\ba =[a_1 \ \cdots \ a_K]^\rT$.
For user $k \in \cA$, the following steps are required:
\begin{enumerate}
    \item User $k$ sends $L_k$ to the BS (via uplink transmission).
\item The BS performs the task with $(L_k, B_k)$ using its CPU (i.e., the server CPU) and obtains the outcome of the task, which is denoted by $Y_k$
(this is also the size of the outcome).
\item The outcome, $Y_k$, is sent to user $k$  (via downlink transmission).  
\end{enumerate}
On the other hand, for user $k \notin \cA$, the following steps are required:
\begin{enumerate}
\item The BS sends the server data, $B_k$, to user $k$ (via downlink transmission).
\item User $k$
performs the task with $(L_k, B_k)$ using its CPU (i.e., the local CPU) and obtains the outcome of the task.  
\end{enumerate}
Note that an offloading user needs to have both uplink and downlink transmissions, while a non-offloading user needs to have downlink transmission as illustrated in Fig.~\ref{Fig:Fig_DS}.

\subsection{Communication Model}

Throughout the paper, we assume dynamic time division duplexing (TDD) for uplink and downlink transmissions with a time slot that can be flexibly divided.

For uplink transmissions, simultaneous transmissions by multiple offloading users are allowed \cite{Chen16}. Let $h_{k}$ denote the channel coefficient from user $k$ to the BS. 
If $k \in \cA$, the uplink transmission rate becomes
\be 
u_k (\ba) = W \log_2 \left( 1 + \frac{P_k \beta_k}{
\sum_{i \in \cA \setminus k} P_i \beta_i + N_0}
\right),
    \label{EQ:UT}
\ee 
where $W$ is the system bandwidth, $P_k$ is the transmit power of user $k$, $\beta_k = |h_k|^2$, and $N_0$ is the variance of the background noise at the BS. 

For downlink transmissions, the BS uses time division multiple access (TDMA) to send data to the offloading users. The data rate of downlink transmission to user $k$ is given by
\be 
v_k = W \log_2 \left(1 + 
\frac{\bar P_k \beta_k}{N_0}
\right) ,
    \label{EQ:DT}
\ee 
where $\bar P_k$ is the transmit power of the BS to user $k$. In \eqref{EQ:DT},
the channel power gain, $\beta_k$, is assumed to be the same as that for uplink transmission because of the channel reciprocity of TDD.
As shown in \eqref{EQ:DT}, each user only needs to perform single-user decoding as there is no interference, while the BS needs to perform multiuser decoding for all offloading users as in \eqref{EQ:UT}. 

Note that for uplink transmissions, TDMA can also be used as in \cite{You17}, and a high transmission rate can be achieved with a 
high transmit power, $P_k$. On the other hand, when simultaneous transmissions are considered, due to the interference, the uplink transmission rate in \eqref{EQ:UT_1} becomes limited, while it would not be necessary to have a high transmit power, $P_k$. 
Thus, simultaneous transmissions could be desirable for users of limited transmit power.

From \eqref{EQ:UT} and \eqref{EQ:DT},
the downlink transmission time of user $k$ is given by
\begin{align}
\bar T_k & = \left\{
\begin{array}{ll}
\frac{Y_k}{v_k}, & \mbox{if $k \in \cA$} \cr
  \frac{B_k}{v_k}, & \mbox{if $k \notin \cA$}. 
\end{array}
\right. 
\end{align}
For the offloading users $k \in \cA$, the uplink transmission  time becomes
$T_k = \frac{L_k}{u_k (\ba)}$.
Since all the offloading users transmit simultaneously, the total uplink transmission time becomes
\be 
T_{\rm up} = \max_k T_k = \max_k \frac{L_k}{u_k (\ba)}.
    \label{EQ:Tup}
\ee 
As mentioned earlier, 
the total transmission time of an offloading user includes both uplink and downlink transmission time. On the other hand, the total transmission time of a non-offloading user only includes downlink transmission time. 

Note that 
when uploading is complete, no further transmission is required.
Thus, some offloading users may end up transmitting and the interference is reduced in the later part of the uplink transmission. As a result, $u_k (\ba)$ may be larger, meaning that $T_{\rm up}$ in \eqref{EQ:Tup} can be seen as an upper-bound.


\subsection{Energy Model}

Let $g_0$ be the consumed energy per CPU cycle at the server and 
$g_k$ at user $k$. Thus, when the task of user $k$ is locally performed, the consumed energy to perform the task becomes $g_k C_k$. On the other hand, when offloading, the consumed energy at the server becomes $g_0 C_k$.
It is also necessary to include the energy for transmissions of data. For user $k \in \cA$, the total consumed energy is given by
\be 
E_k = g_0 C_k + P_k \frac{L_k}{u_k (\ba)} + \bar P_k  \frac{Y_k}{v_k},
\ee
where the first term is the energy for performing the computation at the server, the second term is that for uplink transmission, and the last term is that for downlink transmission. Similarly, the total consumed energy for user $k \notin \cA$ can be found. As a result, we have
\begin{align}
E_k & = \left\{
\begin{array}{ll}
g_0 C_k + P_k \frac{L_k}{u_k (\ba)} + \bar P_k  \frac{Y_k}{v_k}, & \mbox{if $k \in \cA$} \cr
g_k C_k + \bar P_k \frac{B_k}{v_k}, & \mbox{if $k \notin \cA$}. 
\end{array}
\right. 
    \label{EQ:Ek}
\end{align}

\section{Minimization of Total Energy with Channel Inversion Power Control} 

In this subsection, we consider an optimization problem to minimize the total consumed energy with channel inversion power control policy.

\subsection{Channel Inversion Power Control}

As TDD is assumed, signal processing techniques can be used to estimate the uplink channel at a user from the pilot signal transmitted by the BS thanks to the channel reciprocity. Then,
the channel inversion power control policy for uplink transmission is used as follows:
\be 
P_k = \frac{P_{\rm BS}}{\beta_k} \ \mbox{and} \ \bar P_k = \frac{P_{\rm user}}{\beta_k} ,
\ee 
where $P_{\rm BS}$ and $P_{\rm user}$ represent the effective received signal powers at the BS and a user, respectively. Since the maximum transmit power is limited, when $\beta_k$ is too small (due to deep fading), $P_k$ or $\bar P_k$ can be higher than the maximum transmit power and has to be truncated. However, for convenience, we do not consider any truncation and assume that the users of deep fading (i.e., $\beta_k \le \epsilon$ for some $\epsilon > 0$) are excluded.

\subsection{Minimizing Energy Subject To Transmission Time}

In addition to the channel inversion power control policy,
we assume that in order to support all $K$ users, a time slot of length $\tau$, is allocated. Then, it is necessary that the total transmission time of all $K$ users for uplink and downlink transmissions is limited, i.e., $\sum_{k=1}^K \bar T_k + T_{\rm up} \le \tau$. Here, we assume that the computation time at the server is sufficiently short and ignored in the total transmission time constraint.
We can formulate the following problem to minimize the total consumed energy subject to the total transmission time constraint:
\begin{align}
\hat \ba = \argmin_{\ba \in \uB^K} \sum_{k=1}^K E_k 
\ \ \mbox{subject to (s.t.)} \ \sum_{k=1}^K \bar T_k +  T_{\rm up} \le \tau,
    \label{EQ:O1}
\end{align}
where $\uB = \{0,1\}$.
An exhaustive search requires a complexity of order $2^K$.
Thus, for a large $K$, it is necessary to find an approach that can find the solution
with a low complexity.

Note that the objective function in \eqref{EQ:O1} is the total consumed energy at the BS (server) as well as $K$ users. In \cite{You17}, the objective function is the total energy consumed only by  users. On the other hand, in \cite{Chen16}, the energy consumed for signal transmissions is not included in the objective function. Since we consider the case that tasks need to use datasets at storage servers, it is necessary to take into account the cost to send datasets by the BS to non-offloading users, and the objective function has to be the total consumed energy as in \eqref{EQ:O1}.


\subsection{Finding the Solution}   \label{SS:Sol}

In this subsection, we show that the problem in \eqref{EQ:O1} can be solved by converting it into multiple binary integer linear programming problems. Each problem can be solved by a standard technique such as the branch and bound method \cite{Papadimitriou98}.

In order to find the solution of \eqref{EQ:O1},
we assume that the number of the offloading users is $n$
(i.e., $|\cA|= n$). 
Due to the channel inversion power control, from \eqref{EQ:UT}, we can show that
\begin{align} 
u_k (\ba) = u(n) = W \log_2 \left( 1 + \frac{\gamma_{\rm BS}}{
(n-1) \gamma_{\rm BS} + 1}
\right), \forall k
    \label{EQ:UT_1}
\end{align}
where $\gamma_{\rm BS} = \frac{P_{\rm BS}}{N_0}$ is the 
signal-to-noise ratio (SNR) at the BS (for uplink transmission), and
$v_k = v = W \log_2 (1+ \gamma_{\rm user})$, 
where $\gamma_{\rm user} = 
\frac{P_{\rm user}}{N_0}$ is the SNR at a user (for downlink transmission).
As a result, for $n \in \{1,\ldots, K\}$, we have
\begin{align}
T(n) = T_{\rm up} + \sum_{k=1}^K \bar T_k 
= T_0 + \frac{L_{\rm max}}{u(n)} +\sum_{k=1}^K
\frac{Y_k-B_k}{v}a_k, 
    \label{EQ:Tn}
\end{align}
where $L_{\rm max} = \max_k L_k$ and 
$T_0 = \frac{\sum_{k=1}^K B_k}{v}$  is the total transmission time when there is no offloading user (i.e., $T(0) = T_0$). 
Similarly, from \eqref{EQ:Ek}, we can show that
\begin{align}
E (n) = \sum_{k=1}^K E_k 
= E_0 + \sum_{k=1}^K e_k (n) a_k,
    \label{EQ:En}
\end{align}
where 
$E_0 = \sum_{k=1}^K g_k C_k + \frac{P_{\rm user} B_k}{\beta_k v}$ and 
$e_k (n) = (g_0 - g_k) C_k +
\frac{P_{\rm BS} L_k}{\beta_k u(n)}  +
\frac{P_{\rm user} (Y_k - B_k)}{\beta_k v}$.
Then, letting 
$d_k = \frac{Y_k-B_k}{v}$, for a given $n$, the optimization problem becomes a binary integer linear programming as follows:
\begin{align}
\hat \ba (n) =   \argmin_{\ba \in \uB^K} E_0 + \bee^\rT \ba 
\ \ \mbox{s.t.} \
\left\{
\begin{array}{l}
\bd^\rT \ba \le \tau - T_0 (n)\cr  
\bone^\rT \ba = n, \cr 
\end{array}
\right.
    \label{EQ:O2}
\end{align}
where $\bee = [e_1 (n) \ \cdots \ e_K (n)]^\rT$,
$\bd = [d_1   \ \cdots \ d_K  ]^\rT$, and 
$\bone = [1 \ \cdots \ 1]^\rT$.
The second constraint is due to the assumption that $|\cA| = n$
or $||\ba||_1 = n$.

We can solve the problem in \eqref{EQ:O2} for given $n \in \{1, \ldots, K-1\}$. Let $\hat E (n)$ denote the minimum energy, which is finite if there exists a feasible solution. Otherwise, $\hat E(n) = \infty$. Note that if $n = 0$ or $K$, there is no need to solve \eqref{EQ:O2} as $\hat \ba (n)$
is $\b0$ or $\bone$, respectively. Then, the solution of \eqref{EQ:O1} becomes 
\be 
\hat \ba = \hat \ba (n^\ast),
\ee 
where $\hat E(n^*) \le \hat E(n)$, $\forall n \ne n^\ast$. Note that if $\hat E(n^\ast) = \infty$, then there is no feasible solution.
If no feasible solution is available, the length of time slot, $\tau$, can be increased.

\subsection{Mean Transmission Time}

Assume that each of $L_k$, $B_k$, and $C_k$ is  independent and identically distributed (iid).
Furthermore, the output size of user $k$'s task, $Y_k$, is given by
$Y_k = f (L_k + B_k)$, where $f(\cdot)$ is a certain function, i.e., the size of the task output is decided by the total input size, $L_k +B_k$. For simplicity, we consider the following linearization of $f(x)$:
$f(x) \approx c_0 + c_1 x$, where $c_0 > 0$ is the length of the output that is independent of the input and $c_1 > 0$ is a constant, which is assumed to be small. For example, suppose that the task is about classification (where $L_k$ is a test image and $B_k$ is a set of images for comparison). Then, 
the output becomes a label, which requires a few bits to represent. 
Then, from \eqref{EQ:Tn}, we can show that 
the average total transmission time becomes
\begin{align}
\uE[T(n)] 
& = \frac{K \bar B}{v} +
\frac{\uE[L_{\rm max}]}{u(n)} + n \frac{\uE[f(L_k+B_k)] - \bar B}{v} \cr 
& \approx \frac{K \bar B}{v} +
\frac{\uE[L_{\rm max}]}{u(n)} + n \frac{c_0 + c_1 \bar L - (1-c_1) \bar B}{v}   \cr 
&  \approx \frac{K \bar B}{v} + n \theta,
\end{align} 
where 
$ \theta = \frac{\uE[L_{\rm max}] \ln 2}{W} + 
\frac{ c_0 + c_1 \bar L - (1- c_1) \bar B}{v}$.
The last approximation is due to $u(n) \approx 
W \log_2 \left(1 + \frac{1}{n} \right) \approx \frac{W}{n \ln 2}$. Thus, if $\theta > 0$, we can see that $\uE[T(n)]$ increases with $n$, which means that the number of offloading users tends to be small to meet the constraint for a given $\tau$.
We can also find the maximum number of offloading users for a given 
$\tau$, denoted by $\bar n$, which helps reduce the complexity to find $\hat \ba$ in Subsection~\ref{SS:Sol} as \eqref{EQ:O2} is to be solved for $n \in \{1,\ldots, \bar n\}$.
On the the hand, for $\theta$ close to $0$ or less than 0,  the total  transmission time and number of offloading users, $n$, become more or less independent. In this case, with $\tau > \frac{K \bar B}{v}$, the offloading users can be decided to minimize the total consumed energy while being less affected by the total transmission time constraint.

\section{Simulation Results}

For simulations, 
we assume that $L_k$, $B_k$, and $C_k$ are independent, while  
$L_k \sim {\rm Exp} (\bar L)$, 
$B_k  \sim {\rm Exp} (\bar B)$, and
$C_k  \sim {\rm Exp} (\bar C)$. 
In addition, let  $f(x) = c_1 x$, where $c_1 = 0.1$. 
It is assumed that the unit of $L_k$, $B_k$, and $C_k$ is normalized by $W$. Thus, the unit becomes the number of bits per Hz.
We assume that $g_0$ is normalized (i.e., $g_0 = 1$), while $g_k \sim 
{\rm Unif}[0, \bar g]$, where $\bar g = 10$. For the channel gains $\beta_k$, we assume a modified Rayleigh fading such that $\beta_k$ follows a shifted exponential distribution: 
$\beta_k \sim \zeta e^{-\zeta(\beta_k - \epsilon)}$, 
where $\beta_k \ge \epsilon= 0.05$ and $\zeta = \frac{1}{1-\epsilon}$.

\begin{figure}[thb]
\begin{center}
\includegraphics[width=\figwidth]{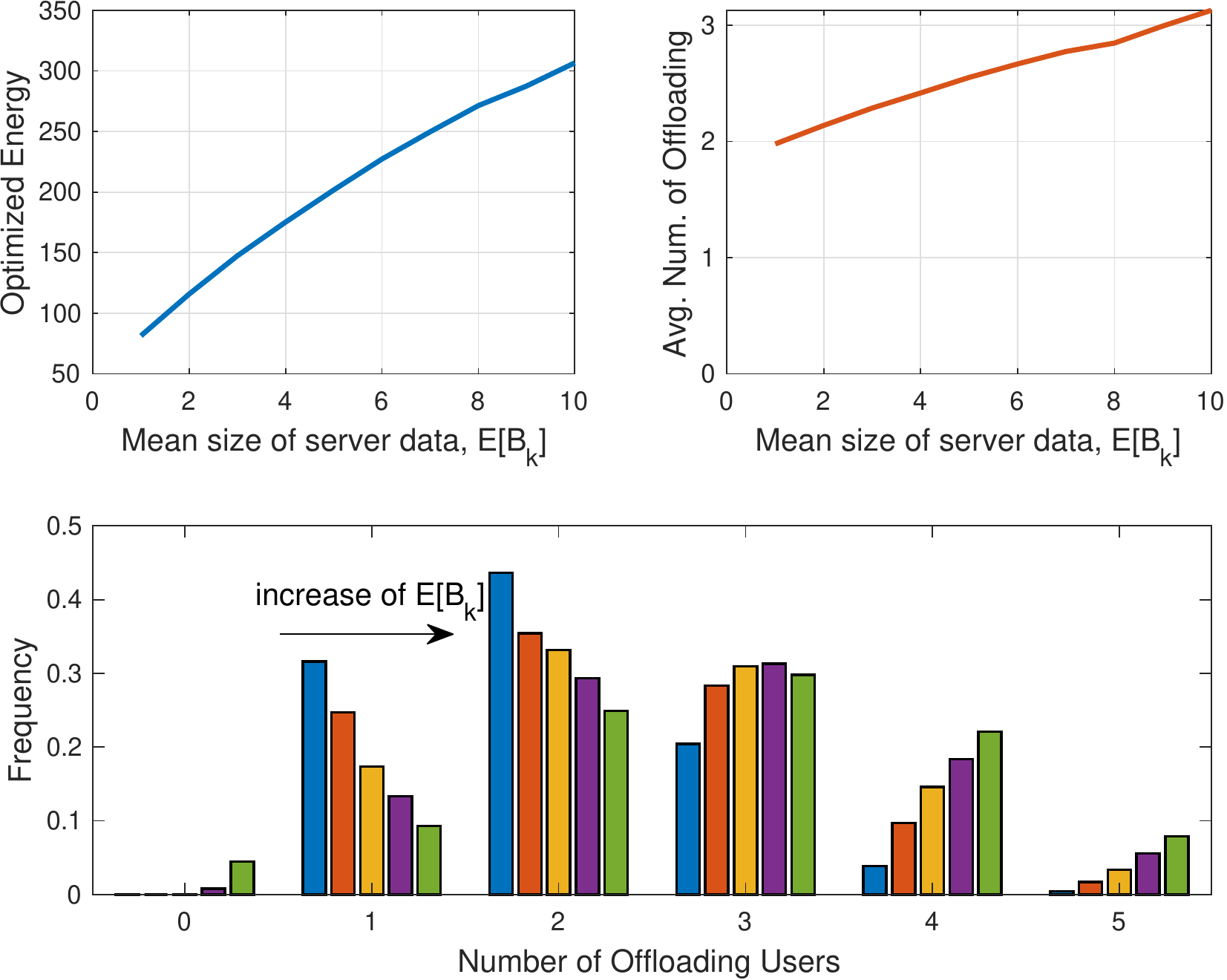} 
\end{center}
\caption{Performance for different size of server data, $\bar B$,
when $K = 10$, $\gamma_{\rm BS} = 3$,
$\gamma_{\rm user} = 6$, $\bar C = 1$, $\bar L = 2$, and $\tau = 
\frac{10 K}{\log_2 (1+ \gamma_{\rm user})} = 
35.63$.}
        \label{Fig:plt1}
\end{figure}

In Fig.~\ref{Fig:plt1}, with
$K = 10$, $\gamma_{\rm BS} = 3$,
$\gamma_{\rm user} = 6$, $\bar C = 1$, $\bar L = 2$,  and $\tau =  T_0 \bigl|_{\bar B = 10} = \frac{10 K}{\log_2 (1+ \gamma_{\rm user})} = 
35.63$, the performance is shown when 
the mean size of server data, $\bar B$, varies. The total consumed energy (shown in the top-left subplot)
and average number of offloading users (shown in the top-right subplot) increase with $\bar B$. This results from the fact that the cost of downloading server data increases with $\bar B$ (which can lead to more offloading users). 
This behavior can  also be seen in the empirical distribution of the number of offloading users (shown in the bottom subplot).
It is also shown that when $\bar B \le \bar L = 2$, the probability that there are 2 offloading users is the highest.  However, as $\bar B$ increases, the probability that there are more than 3 offloading users increases.

\begin{figure}[thb]
\begin{center}
\includegraphics[width=\figwidth]{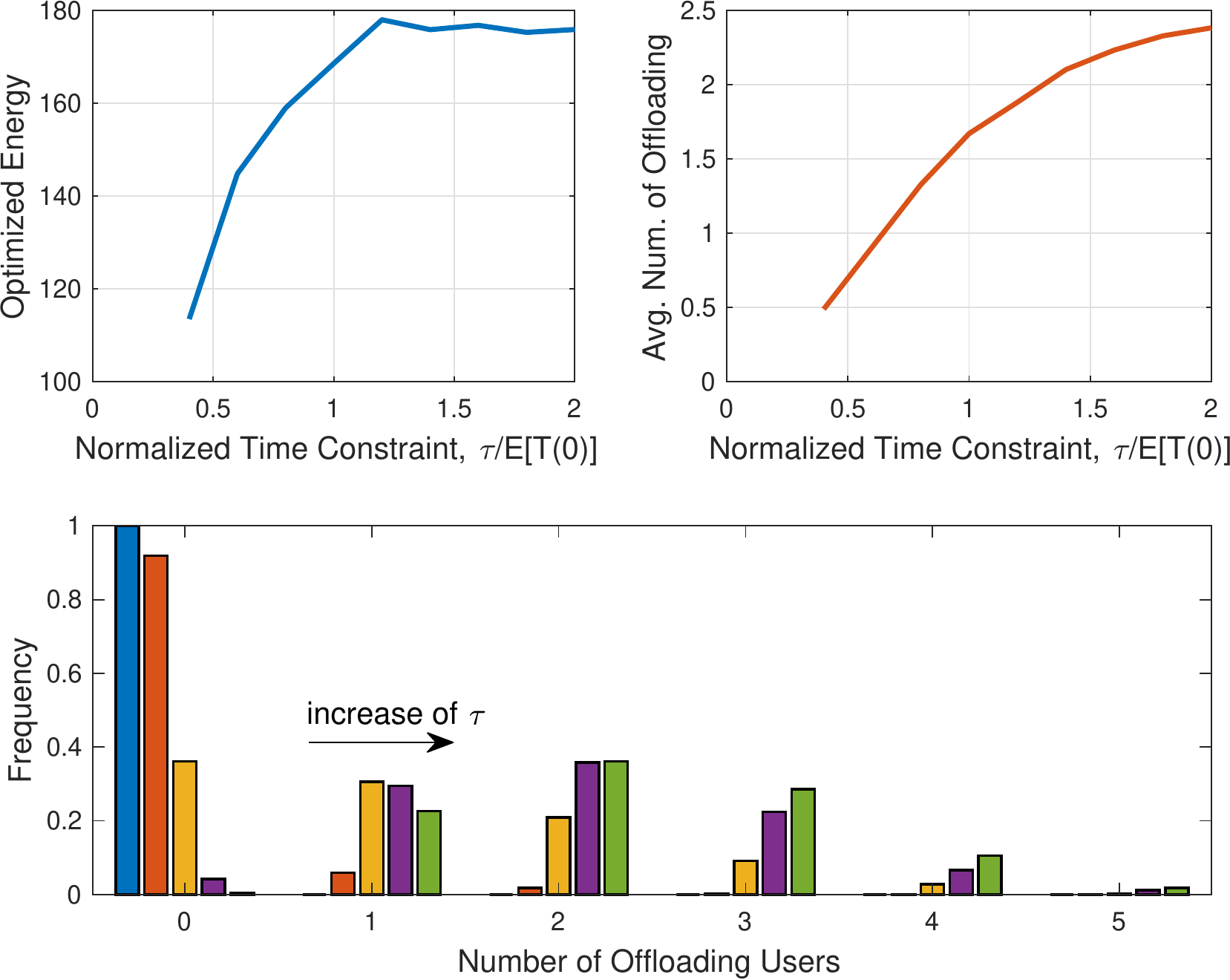} 
\end{center}
\caption{Performance for different threshold of total transmission time, $\tau$,
when $K = 10$, $\gamma_{\rm BS} = 3$,
$\gamma_{\rm user} = 6$, $\bar C = 1$, $\bar L = 2$,  and $\bar B = 4$.}
        \label{Fig:plt2}
\end{figure}

Fig.~\ref{Fig:plt2} shows the
performance for different threshold of total transmission time, $\tau$, when $K = 10$, $\gamma_{\rm BS} = 3$,
$\gamma_{\rm user} = 6$, $\bar C = 1$, $\bar L = 2$,  and $\bar B = 4$.
As $\tau$ increases (i.e., the constraint is more relaxed), the total consumed energy increases (shown in the top-left subplot) and there are more offloading users (shown in the top-right and bottom subplots).

\section{Concluding Remarks}

We considered computation offloading when users need to use data at cloud storage. A minimization problem of the consumed energy subject to a total transmission time constraint was formulated and its solution was obtained via binary integer linear programming formulation. For further research, we will extend the approach to take into account the  cost of retrieving data from cloud storage servers.

\bibliographystyle{ieeetr}
\bibliography{ref}

\begin{thebibliography}{10}

\bibitem{Yang08}
K.~Yang, S.~Ou, and H.-H. Chen, ``On effective offloading services for
  resource-constrained mobile devices running heavier mobile internet
  applications,'' {\em IEEE Communications Magazine}, vol.~46, no.~1,
  pp.~56--63, 2008.

\bibitem{Kumar10}
K.~Kumar and Y.-H. Lu, ``Cloud computing for mobile users: Can offloading
  computation save energy?,'' {\em Computer}, vol.~43, no.~4, pp.~51--56, 2010.

\bibitem{Barbarossa14}
S.~Barbarossa, S.~Sardellitti, and P.~Di~Lorenzo, ``Communicating while
  computing: Distributed mobile cloud computing over {5G} heterogeneous
  networks,'' {\em IEEE Signal Processing Magazine}, vol.~31, no.~6,
  pp.~45--55, 2014.

\bibitem{Mach17}
P.~Mach and Z.~Becvar, ``Mobile edge computing: A survey on architecture and
  computation offloading,'' {\em IEEE Communications Surveys Tutorials},
  vol.~19, no.~3, pp.~1628--1656, 2017.

\bibitem{Abbas18}
N.~Abbas, Y.~Zhang, A.~Taherkordi, and T.~Skeie, ``Mobile edge computing: A
  survey,'' {\em IEEE Internet of Things Journal}, vol.~5, no.~1, pp.~450--465,
  2018.

\bibitem{Chen16}
X.~Chen, L.~Jiao, W.~Li, and X.~Fu, ``Efficient multi-user computation
  offloading for mobile-edge cloud computing,'' {\em IEEE/ACM Transactions on
  Networking}, vol.~24, no.~5, pp.~2795--2808, 2016.

\bibitem{You17}
C.~You, K.~Huang, H.~Chae, and B.-H. Kim, ``Energy-efficient resource
  allocation for mobile-edge computation offloading,'' {\em IEEE Transactions
  on Wireless Communications}, vol.~16, no.~3, pp.~1397--1411, 2017.

\bibitem{Zhou21}
P.~Zhou, K.~Shen, N.~Kumar, Y.~Zhang, M.~M. Hassan, and K.~Hwang,
  ``Communication-efficient offloading for mobile-edge computing in {5G}
  heterogeneous networks,'' {\em IEEE Internet of Things Journal}, vol.~8,
  no.~13, pp.~10237--10247, 2021.

\bibitem{Anajemba20}
J.~H. Anajemba, T.~Yue, C.~Iwendi, M.~Alenezi, and M.~Mittal, ``Optimal
  cooperative offloading scheme for energy efficient multi-access edge
  computation,'' {\em IEEE Access}, vol.~8, pp.~53931--53941, 2020.

\bibitem{Choi22_OL}
J.~Choi, ``Random-access-based multiuser computation offloading for devices in
  {IoT} applications,'' {\em IEEE Internet of Things Journal}, vol.~9, no.~21,
  pp.~22034--22043, 2022.

\bibitem{BUYYA2009599}
R.~Buyya, C.~S. Yeo, S.~Venugopal, J.~Broberg, and I.~Brandic, ``Cloud
  computing and emerging it platforms: Vision, hype, and reality for delivering
  computing as the 5th utility,'' {\em Future Generation Computer Systems},
  vol.~25, no.~6, pp.~599--616, 2009.

\bibitem{Wu10}
J.~Wu, L.~Ping, X.~Ge, Y.~Wang, and J.~Fu, ``Cloud storage as the
  infrastructure of cloud computing,'' in {\em 2010 International Conference on
  Intelligent Computing and Cognitive Informatics}, pp.~380--383, 2010.

\bibitem{Papadimitriou98}
C.~Papadimitriou and K.~Steiglitz, {\em Combinatorial Optimization: Algorithms
  and Complexity}.
\newblock Dover Books on Computer Science, Dover Publications, 1998.

\end{thebibliography}

\vfill
\end{document}